# Prejudiced Futures? – Algorithmic Bias in Time Series Forecasting and Its Ethical Implications

Bagattini Alexander*, Chen Shao

a.bagattini@kit.edu, chen.shao2@kit.edu

**Abstract**

Time series prediction algorithms are increasingly central to decision-making in high-stakes domains such as healthcare, energy management, and economic planning. Yet, these systems often inherit and amplify biases embedded in historical data, flawed problem specifications, and socio-technical design decisions. This paper critically examines the ethical foundations and mitigation strategies for algorithmic bias in time series prediction. Drawing from interdisciplinary literature in philosophy, computer science, and law, we argue that bias is not merely a technical anomaly but a systemic issue permeating all stages of the algorithmic pipeline, from problem formulation and data collection to modeling choices and deployment. We outline how predictive models, particularly in temporally dynamic domains, can reproduce structural inequalities and emergent discrimination through proxy variables and feedback loops.

The paper advances a threefold contribution: First, it reframes algorithmic bias as a socio-technical phenomenon rooted in normative choices and institutional constraints. Second, it offers a structured diagnosis of bias sources across the pipeline, emphasizing the need for causal modeling, interpretable systems, and inclusive design practices. Third, it advocates for structural reforms that embed fairness through participatory governance, stakeholder engagement, and legally enforceable safeguards. Special attention is given to fairness validation in dynamic environments, proposing multi-metric, temporally-aware, and context-sensitive evaluation methods. Ultimately, we call for an integrated ethics-by-design approach that positions fairness not as a trade-off against performance, but as a co-requisite of responsible innovation. This framework is essential to developing predictive systems that are not only effective and adaptive but also aligned with democratic values and social equity.

*corresponding author.

**Introduction**

Recent years have witnessed growing concern over the ethical implications of algorithmic decision-making, particularly in high-stakes domains such as healthcare, public policy, and infrastructure management. Much of this concern centers on algorithmic bias, how it arises, how it manifests, and how it reinforces or amplifies existing social inequalities. While the problem of bias in machine learning has received sustained attention in areas such as classification and image recognition, considerably less focus has been given to the unique challenges posed by time series prediction. This paper seeks to address that gap by critically examining how bias operates within the specific technical and ethical landscape of time series modeling.

Time series prediction plays an increasingly central role in domains where temporal data is used to guide consequential decisions. In healthcare, models for time series prediction forecast patient risk and inform clinical interventions. In energy systems, they support load balancing and policy design. In economics, they guide forecasts that shape fiscal and monetary policy. Despite their growing influence, these systems often rest on the assumption that historical patterns can be extrapolated into the future, an assumption that fails to account for social change, intervention effects, or structural inequalities embedded in the past. As a result, time series models are vulnerable to biases that are not just technical errors, but normative failures with far-reaching consequences.

This paper's central thesis is that algorithmic bias in time series prediction is best understood as a systemic phenomenon rooted in normative choices made throughout the model development pipeline. These choices span problem specification, data collection, model selection, and deployment. Importantly, bias in this context is not confined to the inclusion of sensitive features like race or gender; it can also emerge through proxy variables, feedback loops, and institutional defaults that silently encode unequal social realities. While these risks are shared with other areas of machine learning, time series models add specific challenges: temporal dependencies, non-stationarity, and evolving environments complicate both diagnosis and mitigation of bias. Moreover, because predictions influence the very future they are meant to anticipate, time series systems can initiate feedback cycles that deepen existing disparities over time.

To navigate these complexities, this paper draws from an interdisciplinary foundation that integrates perspectives from philosophy and computer science, and law. Philosophical inquiry enables us to clarify what is meant by fairness and accountability in predictive settings, and to critique the normative assumptions behind formal modeling choices. Computational methods offer tools for bias detection, causal inference, and interpretable modeling. Together, these perspectives support a more comprehensive and context-sensitive approach to fairness in time series prediction.

Our contribution is threefold: *Ethical Reframing of Bias in Time Series Prediction:* We offer a conceptual framework that understands algorithmic bias not merely as a deviation from accuracy, but as a misalignment with ethical and democratic values. Drawing from philosophical and legal literature, we explore how normative decisions are embedded in the translation of social goals into predictive tasks. *Pipeline-Level Diagnosis and Causal Accountability*: We provide a structured analysis of how bias arises at each stage of the predictive modeling pipeline. Particular attention is given to the challenges posed by proxy variables, concept drift, and the unintended consequences of deploying models in dynamic environments. We argue that causal modeling and interpretability are key to meaningful diagnosis and accountability. *Institutional and Procedural Solutions*: Finally, we advocate for reforms that go beyond technical fixes, including participatory design, stakeholder engagement, interdisciplinary education, and regulatory oversight. Fairness, we argue, must be built into the system through procedural safeguards, not added retroactively as a constraint.

By focusing on time series prediction, this paper extends the current debate on algorithmic fairness into a domain that has been relatively underexplored despite its growing real-world impact. Our aim is not only to identify and mitigate bias but to show how ethical reasoning can, and must, inform the design and deployment of predictive technologies from the outset. In high-stakes, temporally sensitive domains, fairness is not an optional ideal but a practical requirement for responsible innovation.

## 1. Conceptualizing Algorithmic Bias in Time Series Prediction

Algorithmic bias refers to systematic and unfair deviations in the behavior or outcomes of algorithms, particularly those used in decision-making processes that affect individuals and groups. This phenomenon becomes especially critical when algorithms are deployed in high-stakes domains such as healthcare, criminal justice, finance, or education. At its most general, algorithmic bias can be understood as a systematic deviation from a normative standard, be it statistical accuracy, moral fairness, or legal neutrality. (Bryson 2020, Coeckelbergh 2022) An algorithm may be statistically biased if its predictions diverge significantly from reality, or morally biased if it disadvantages individuals based on illegitimate features like race or gender (Fazelpour & Danks, 2021). Importantly, bias is not merely a technical flaw. It often stems from design decisions, dataset limitations, and social contexts. Fazelpour and Danks emphasize that bias can emerge at any stage of the development process of algorithm (what they call „algorithmic pipeline"): during problem formulation, data collection, model training, or deployment. (ibid., p. 4ff.) This refers to their argument that converting inherently complex, value-driven objectives into quantifiable targets is not merely technical but involves normative decisions. According to their approach, such operationalization reflects social values and institutional priorities, potentially introducing biases by excluding outcomes significant to marginalized groups. Hence, they emphasize the necessity of explicitly addressing these normative choices to prevent systematic bias. (Ibid., p. 2ff)

While these concerns are well known and apply broadly, they take on particular significance in the context of time series prediction. Time series models, used to forecast sequential phenomena such as disease spread, energy demand, or market trends, bring with them unique methodological and ethical challenges. These models often rely on assumptions of temporal correlation and causal continuity, that is, that patterns in the past will persist into the future. However, this assumption can be problematic in real-world contexts marked by social change or intervention. For example, in healthcare, models predicting hospital admissions or disease outbreaks may be trained on data that reflect entrenched disparities in care access. During the COVID-19 pandemic, algorithms used to allocate ventilators or hospital beds often relied on past medical expenditure as a proxy for healthcare need, thus reinforcing the historical underinvestment in Black patients (Obermeyer et al., 2019; Giovanola & Tiribelli, 2022). In energy forecasting, similar concerns emerge. Predictive models used for electricity demand management or renewable integration are often trained on consumption data that reflect socio-economic stratification. For instance, lower-income households may exhibit different usage patterns due to energy poverty, underrepresentation in smart grid data, or restricted access to efficient appliances. If these groups are not adequately captured in the training data, models may underpredict their needs, affecting grid allocation or policy targeting (Liu et al., 2020; Hesse et al., 2019). Moreover, energy data is highly susceptible to seasonal and structural non-stationarities, factors like weather anomalies, policy changes, or energy market shocks can disrupt temporal regularities, causing models to make biased or inaccurate predictions.

A related concern is the use of proxy variables and latent correlations in time series data. As Matzner and Mann argue, algorithmic profiling can create new forms of emergent discrimination, even in the absence of explicit demographic attributes. (Matzner & Mann 2019) Time series models may infer group membership indirectly through features like zip codes or usage frequencies, data points that correlate with socio-economic status or race, resulting in discriminatory impacts even when protected categories are formally excluded. Addressing bias in time series prediction thus requires a multi-disciplinary approach. Researchers in philosophy, computer science, and the social sciences increasingly argue for a causal perspective on prediction, emphasizing the need to disentangle correlation from causation, especially in policy-relevant applications (Pearl & Mackenzie, 2018; Barabas et al., 2018). Transparency, stakeholder engagement, and context-sensitive evaluation metrics are essential to ensure that predictive models advance both accuracy and fairness, rather than unintentionally reinforcing historical inequities.

Our contribution includes: 1) *Problem Reframing through Ethical Foundations*: We conceptualize algorithmic bias in time series prediction as a systemic and socio-technical phenomenon, drawing from interdisciplinary literature in computer science, philosophy, and social theory. 2) *Pipeline-Level Bias Attribution*: We provide a structured analysis of how different stages, problem formulation, data collection, model design, and deployment, contribute to bias propagation. Based on the insights, we highlight the disconnect between ethical intent and practical implementation. Our analysis reveals that practitioners often lack the conceptual tools or institutional support needed to realize fairness-aware systems, especially under commercial or performance-driven constraints. 3) *Call for Structural Reforms in ML Practice*: Beyond technical interventions, we argue for systemic changes, such as curricular reform, interdisciplinary collaboration, and regulatory oversight, as necessary conditions for making fairness operational at scale. We advocate for an integrated view of fairness that positions ethical and technical excellence as co-dependent rather than competing objectives.

## 2. Sources and Mechanisms of Bias

An important first step for orientation when considering the ethical and social implications of algorithms is a more precise understanding of how biases arise and how they should be evaluated. In their comprehensive account of algorithmic bias, Fazelpour and Danks present a nuanced understanding of the *sources and mechanisms of bias* that pervade predictive models. Their analysis underscores the idea that algorithmic bias is not merely a byproduct of flawed code or technical oversight but rather a systemic issue embedded in each stage of the algorithmic pipeline, from problem specification and data collection to model training and deployment. At its core, algorithmic bias reflects a mismatch between the normative ideals we expect from algorithmic decisions and the value-laden practices that inform their development.

One of the important issues raised by Fazelpour and Danks is that bias often originates in the problem specification phase. (Fazelpour and Danks 2021, p. 4) In theory, developers are tasked with translating complex, value-laden goals, such as 'student success' or 'recidivism risk', into precise, quantifiable targets for prediction, a process that requires not only technical skill but also normative judgment, as it involves deciding which aspects of these multifaceted goals are prioritized, simplified, or excluded altogether. (This target-setting is a key step in problem formulation, where complex social goals must be operationalized into measurable variables, inevitably reflecting normative choices.) However, in practice, the authority to define such benchmarks often lies with a few highly influential institutions. These entities shape not only technical standards but also the normative assumptions embedded in predictive models, raising important questions about accountability, inclusivity, and representational legitimacy. This process, though seemingly neutral, inevitably involves normative choices. For example, deciding whether student success is measured by GPA (grade point average), post-graduation salary, or first-year retention rates is not a purely technical matter but one shaped by institutional priorities and social values. Different definitions can disproportionately affect marginalized groups, resulting in what Fazelpour and Danks term *omitted payoff bias*, a form of bias that emerges when key outcomes or stakeholder interests are excluded from consideration. (ibid., p. 8)

Bias is further perpetuated in the data collection stage, where two primary mechanisms come into play: historical biases embedded in real-world systems, and limitations in data measurement. When algorithms are trained on data derived from unjust systems, say, healthcare datasets reflecting systemic racial disparities in treatment access, the models will „learn" and replicate these injustices. This phenomenon, colloquially summarized as „bias in, bias out," underscores the moral hazard of relying on historical data without critical scrutiny. Compounding this issue, data collection methods may themselves be biased. For example, surveys distributed only at certain campus locations may fail to capture the needs of less-visible student populations, leading to *sample bias* and poorer model performance for underrepresented groups.

Fazelpour and Danks also highlight how biases in modeling and validation arise through the selection of optimization criteria. Algorithms are not designed in a vacuum; they are built to succeed according to specific metrics, accuracy, precision, F1 score, that implicitly prioritize some outcomes over others. Choosing a metric is itself a value judgment, one that may privilege overall accuracy at the expense of fairness across demographic groups. These trade-offs become particularly salient when fairness constraints, such as equal false-positive rates across groups, conflict with performance goals. The authors point out that the technical impossibility of satisfying all fairness criteria simultaneously forces developers to make normative decisions about which values to uphold. Furthermore, biases in deployment reveal how algorithms can misalign with user values or fail to adapt to new contexts. Even a model that performs well in one environment may produce biased outcomes when transferred elsewhere, due to changes in demographic makeup, institutional norms, or user expectations. Moreover, users may misinterpret algorithmic outputs, especially when models lack transparency or causal interpretability. For instance, using a predictive model trained to identify students at risk of dropping out to make financial aid decisions might inadvertently penalize those who could benefit most from support.

Fazelpour and Danks argue that these mechanisms of bias are not isolated, but often interact in several ways. A biased problem specification can lead to biased data collection, which in turn leads to biased modeling choices and ultimately biased decisions in practice. The solution, then, cannot be purely technical. Instead, they advocate for a process-oriented ethical evaluation that scrutinizes each stage of algorithm development, and which embraces transparency, interdisciplinary collaboration, and stakeholder engagement. By revealing the interdependence of technical and ethical factors, their account offers a compelling roadmap for diagnosing and mitigating algorithmic bias in a principled and context-sensitive manner. Our own approach in the fourth section draws heavily on this model by Fazelnpur and Danks. Before we move on to examining strategies for avoiding biases in more detail in the fourth section, we would like to first discuss some of the ethical and social implications of biases in time series prediction in the following section.

### 3. (Some) Ethical Implications

There are, of course, various ethical issues that are relevant here. We will focus on three below: the disproportinate impact on marginalized groups, erosion of trust and accountability and implications for justice, fairness, and respect for individuals.

One of the most troubling consequences of algorithmic bias in time series prediction is its *disproportionate impact on marginalized communities*. Although time series models are often seen as neutral tools for forecasting future trends, their design, training, and deployment frequently entrench historical inequalities and reproduce forms of systemic discrimination. As highlighted by Fazelpour and Danks, algorithmic bias is not simply a computational flaw, but a value-laden artifact of the social, political, and economic systems in which algorithms are embedded. These systems already disadvantage certain populations, such as racial minorities, women, people with disabilities, and economically disenfranchised groups, and algorithms often exacerbate these inequities under the guise of predictive efficiency. (Fazelnpour & Danks 2021, p. 3ff.) Matzner and Mann, for example, demonstrate that algorithmic profiling, even when stripped of explicit demographic indicators, can recreate discriminatory outcomes through the use of proxy variables. (Mazner & Mann 2019, p. 2) In time series contexts, such proxies may include ZIP codes, temporal spending patterns, or healthcare usage data, all of which correlate strongly with socio-economic status, race, and other marginalized identities. For instance, an energy consumption forecasting model trained primarily on data from affluent neighborhoods may fail to accurately capture the needs of lower-income households who live with energy poverty, limited appliance usage, or shared meter systems. Such underrepresentation can lead to misallocations of energy resources and policy misdirection that compound social exclusion. Moreover, the compounding nature of bias across the algorithmic pipeline amplifies the harm to these groups. As Fazelpour and Danks point out, decisions made during problem specification, such as selecting proxy targets for the assessment of health risks or of student performances, can inadvertently exclude outcomes that matter most to marginalized individuals. (Fazelnpour and Danks 2021, p.6) When this is combined with biased data collection (e.g., missing data from underrepresented communities) and optimization goals that prioritize overall accuracy rather than equitable outcomes, the result is a model that systematically mispredicts and misserves the very populations most in need of fair treatment.

The impact is particularly stark in public health applications. Time series algorithms used during the COVID-19 pandemic to allocate medical resources like ventilators and intensive care beds often relied on past healthcare expenditures as a proxy for need. As noted in both empirical and philosophical literature, this practice effectively penalized underserved groups who historically received less medical care, not due to lower need, but due to discriminatory barriers to access (Obermeyer et al., 2019; Matzner & Mann, 2019). Here, time series models didn't just reflect inequalities; they reinforced them, directing lifesaving resources away from those historically neglected. Importantly, both Matzner and Mann and Fazelpour and Danks stress that traditional anti-discrimination frameworks often fall short in addressing these emergent and intersectional harms. Discrimination in algorithmic systems does not always map neatly onto protected categories like „race" or „gender". It frequently manifests through complex combinations of attributes, or through entirely new algorithmically generated categories that have no social counterpart. This form of emergent discrimination is invisible to conventional regulatory mechanisms, yet deeply impactful in practice.

Therefore, addressing the disproportionate impact of time series prediction on marginalized groups requires a rethinking of what constitutes fairness in algorithmic decision-making. Beyond technical solutions like debiasing algorithms or re-weighting datasets, what is needed is a systemic shift toward ethical design principles that recognize the historical and structural roots of inequality. Intersectional and post-colonial analyses, as suggested by Matzner and Mann, offer vital lenses through which these new forms of invisibility can be made visible. Ethical algorithm design must be anticipatory, inclusive, and reflexive, attuned to the social positions of those it affects and grounded in participatory processes that give marginalized voices a say in how predictive systems are built and deployed.

As algorithmic systems increasingly mediate critical decisions across domains like healthcare, criminal justice, and energy infrastructure, concerns over *trust and accountability* have taken on renewed urgency. In the context of time series prediction, where future outcomes are inferred from historical patterns, the opacity and complexity of algorithms can undermine both the trust of users and the accountability of institutions deploying them. (Burrell 2016) Drawing on the frameworks provided by Cooper et al. and Heaton et al., we will now explore how trust and accountability can be threatened in data-driven systems, and why addressing these issues is central to the ethical use of time series algorithms. Cooper et al. argue that algorithmic systems can exacerbate long-standing barriers to accountability identified by Helen Nissenbaum, such as the „many hands" problem, the use of the computer as scapegoat, and „ownership without liability." (Cooper et al. 2022 p. 879, Nissenbaum 1996) These challenges are amplified in time series prediction, where models are built through complex pipelines involving multiple actors, from data curators and modelers to institutional decision-makers, making it difficult to determine who is answerable when things go wrong. This diffusion of responsibility can lead to an accountability vacuum, in which no actor fully acknowledges blame, and harmed individuals or communities have limited recourse.

Heaton et al. emphasize the link between accountability and trust, showing how the perceived absence of clear responsibility erodes public confidence in algorithmic systems. (Heaton et al. 2023, p. 5) According to them, trust is not a given; it must be earned through transparency, reliability, and the perception that system designers and deployers are ethically and socially responsive. (Compare furthermore Alaieri & Vellino 2016) When time series models produce harmful or biased outcomes, such as underpredicting resource needs for underrepresented populations or triggering unwarranted interventions, users often do not know whom to hold accountable. Worse still, algorithmic agency is often overstated, encouraging a misleading narrative where systems appear autonomous and thereby absolving human actors of moral responsibility. (Heaton et al. 2023, p. 3 ff.) This erosion of accountability has direct consequences for trust. As Heaton et al. point out, users are less likely to trust systems that obscure human agency or evade blame. Conversely, robust accountability mechanisms, such as algorithmic audits, transparent documentation, and mechanisms for contestation, can strengthen trust by signaling that developers and institutions take responsibility for their tools. Yet, as Cooper et al. note, such mechanisms remain rare, and existing standards of care often fall short in addressing sociotechnical harms like bias or exclusion. Legal frameworks have also struggled to keep pace, particularly with respect to novel harms that do not fit neatly within existing categories of liability. The erosion of trust and accountability is not just a technical or organizational failure; it is a profound ethical concern. Trustworthy systems require more than just technical accuracy, they must be embedded within governance structures that prioritize transparency, contestability, and responsiveness to harm. For time series models to be ethically viable, especially in high-stakes domains, stakeholders must be able to identify who is accountable, for what, and under which circumstances. As Cooper et al. argue, achieving this requires combining moral and relational approaches to accountability, creating frameworks that link blameworthiness with social roles and obligations. Only then can we begin to rebuild the trust that algorithmic systems increasingly demand but rarely deserve.

The increasing deployment of algorithmic systems, particularly in predictive modeling such as time series forecasting, raises furthermore profound *ethical implications for justice, fairness, and respect for individuals*. These systems do not operate in neutral vacuums; rather, as emphasized by Fazelpour and Danks, they are embedded in social contexts, shaped by value-laden decisions at every stage, from data collection to optimization goals. (Fazelnpour & Danks 2021, p. 7) Consequently, algorithmic models can reflect, amplify, or even create new injustices, especially when they encode structural inequalities or privilege certain outcomes over others without adequate justification. Justice, broadly understood, requires treating individuals in a way that is morally justified and socially equitable. However, as Hacker points out, algorithmic decision-making often fails to meet this standard because the legal and institutional frameworks intended to ensure non-discrimination are ill-equipped to address complex forms of algorithmic bias. (Hacker 2022) For example, machine learning systems may engage in „proxy discrimination," where seemingly neutral variables (like geographical location or education history) correlate with protected attributes like race or gender, resulting in systematically unjust outcomes. These effects are even enforced when individuals lack the transparency, access, or legal standing to contest such decisions, undermining both procedural justice and substantive fairness.

Fairness, as elaborated by Giovanola and Tiribelli, cannot be reduced merely to non-discrimination or statistical parity. (Giovanola and Tiribelli 2022) Their philosophical analysis calls for a richer, ethically grounded conception of fairness that encompasses both distributive and socio-relational dimensions. Distributive fairness addresses the allocation of goods or burdens (e.g., healthcare, credit, or opportunities), while relational fairness emphasizes the importance of treating individuals with dignity and acknowledging their unique identities and needs. In the context of time series prediction, especially when used in critical domains like healthcare, criminal justice, or resource allocation, this dual commitment is essential. Predictive models that generalize across populations risk obscuring individual-level nuances and thereby violating the moral imperative to respect persons not merely as data points, but as individuals with context-specific claims. Moreover, these ethical failures often occur not through overt malice but through technical oversights and design choices. Here, Fazelpour and Danks warn that algorithmic bias often originates in the problem specification stage, where goals are operationalized in ways that exclude or undervalue marginalized populations. (Fazelpour & Danks 2021, p. 8) For instance, defining „success" or „risk" in ways that are easier to quantify, but less socially meaningful, can lead to models that optimize for convenience rather than justice. These models may inadvertently reinforce systemic disadvantages, thus eroding fairness under the guise of efficiency or predictive accuracy.

Respect for individuals, a foundational principle in moral philosophy, demands that systems do not reduce people to categories or treat them merely as means to ends. Giovanola and Tiribelli argue that fairness must be reconceptualized to include the ethical demand of respect, not only for persons in general but for particular individuals with distinctive life circumstances. In predictive analytics, this means creating mechanisms for individuals to understand, contest, and influence algorithmic decisions that affect them. Yet, as Hacker notes, access to the data and logic behind algorithmic outputs is often shielded by proprietary protections or technical opacity, making it nearly impossible for individuals to assert their rights or dignity within algorithmic systems. (Hacker 2022)

In light of these challenges, achieving justice, fairness, and respect in algorithmic time series prediction requires more than technical fixes. It demands institutional redesigns, interdisciplinary collaboration, and a commitment to substantial ethical values. Legal instruments like the GDPR, when interpreted alongside anti-discrimination frameworks, may offer some recourse by enforcing transparency and enabling audits. But ultimately, what is required is a shift from treating fairness as a statistical constraint to recognizing it as a normative commitment, one that centers ethical values and ensures that algorithmic tools serve rather than subjugate the people they aim to help.

At this point we would like to add a cautionary note: Algorithm development is not merely a technical pursuit aimed at optimizing performance metrics; it is a structured, iterative process requiring careful design, transparent evaluation, and fair comparison. At the heart of this process lies the principle of scientific integrity, experiments must be designed not to favor a desired outcome, but to fairly test the algorithm's claims against meaningful baselines. A well-grounded algorithm development pipeline begins with a clear definition of the problem and the selection of appropriate benchmarks. These baselines serve as reference points, enabling researchers and readers alike to assess the novelty and effectiveness of a proposed method. Crucially, a baseline must be both relevant and robust, comparing against outdated, poorly implemented, or irrelevant methods undermines the credibility of the results. Strong experimental design must also incorporate context-sensitive metrics that reflect real-world implications, especially in terms of fairness, social impact, and accountability. As algorithmic systems increasingly mediate consequential decisions, it is essential to move beyond conventional metrics like accuracy and efficiency, and toward evaluation frameworks that explicitly capture fairness, transparency, and broader societal effects. An often overlooked aspect of the pipeline is the need to continuously update baselines and assumptions as the field evolves. Using legacy results or static datasets risks producing evaluations that are not reflective of current standards. Similarly, over-reliance on a single well-known implementation can stall innovation, especially if new methods are only compared within a narrow experimental scope. Finally, in cases where a novel problem or perspective is introduced, researchers must thoughtfully construct comparisons to whatever reasonable references are available, even if they are indirect or approximate. The goal remains to produce evidence that is not only scientifically rigorous but also socially meaningful and contextually grounded. In this spirit, algorithm development should be viewed as a dynamic, evidence-driven process, one that advances the field not just through innovation, but through the rigor and transparency of how that innovation is evaluated.

## 4. Comprehensive Mitigation Strategies

Algorithmic biases in time series prediction have been identified throughout Sections 2 and 3 as deeply embedded issues arising from multiple stages of the predictive modeling pipeline. These biases include historical inequalities embedded within training data, methodological shortcomings such as reliance on static correlation assumptions, and ethical pitfalls associated with problem specification and data selection practices. Collectively, these biases result in systematic disadvantages, particularly for marginalized groups, undermining fairness, trust, and accountability in critical decision-making contexts such as healthcare, energy distribution, and economic planning. Addressing these biases necessitates comprehensive solutions that integrate both technical innovations and ethical considerations. Technical strategies involve the development of robust modeling frameworks capable of handling data imbalances and adapting to temporal dynamics. Ethical solutions emphasize transparency, stakeholder involvement, and inclusive design processes, ensuring algorithms are responsive to societal needs and normative fairness criteria. This section outlines detailed mitigation strategies that reflect interdisciplinary insights and provide actionable pathways to reduce bias and enhance the ethical deployment of predictive algorithms.

### 4.1 Technical Fixes: Data Related Solutions and Algorithmic Adjustments

Effective strategies to mitigate biases in time series prediction must begin at the data acquisition and preparation stages, addressing both inherent dataset imbalances and biases introduced during data collection.

*Balanced Dataset Collection:* To address biases originating from historical inequalities embedded within datasets, strategies must actively enhance the representation of marginalized and underrepresented groups. These strategies include targeted oversampling of historically underrepresented groups, intentional sampling in less-visible contexts, and expanded data collection to cover socio-economically diverse populations. As emphasized by Fazelpour and Danks, critical scrutiny of data sources and practices is essential to avoid perpetuating "bias in, bias out" scenarios, ensuring algorithmic systems do not replicate systemic injustices embedded in their training data (Fazelpour & Danks, 2021).

*Imbalanced Graph Correlation Learning:* Real-world graphs, such as social networks, transaction graphs, and biological systems, often exhibit structural imbalance. This refers to uneven distributions in node degrees, subgraph densities, or connection patterns. Traditional graph anomaly detection (GAD) methods typically assume structural homogeneity or balanced patterns, which limits their effectiveness in imbalanced settings. Structural imbalance can either conceal true anomalies or mimic anomalous patterns, posing a unique challenge. Moreover, anomalies tend to occur rarely, the amount of data is small, and systems for this type of machine learning are intrinsically limited from a theoretical perspective.

Recent work has introduced novel approaches to address this issue. Xu et al. propose a method based on mixed entropy minimization, which avoids oversampling and enables robust learning from highly imbalanced graph structures. (Xu et al 2024) Their approach, GraphME, improves minority class representation while maintaining structural consistency across the graph. Similarly, Wen et al. integrate graph-based SMOTE techniques (Synthetic Minority Over-sampling Technique) with transformer embeddings to synthetically generate rare node and edge patterns, particularly useful in fraud detection contexts. (Wen et al. 2024) This combination allows for better representation of minority substructures while preserving global graph semantics.

*Data Augmentation:* Time series forecasting models often suffer when training data is scarce or noisy. Data augmentation addresses this gap by synthetically expanding the training set, which can enhance model generalization and robustness. While data augmentation is standard in computer vision and has been studied for time series classification and anomaly detection, its use in time series forecasting has only recently gained traction. Initial studies indicate that augmentation can improve forecasting accuracy, especially in low-data regimes (Wen et al., 2021).

*Mixup*: Creating synthetic examples by linearly combining pairs of time series (and, for supervised tasks, their targets). Originally developed for image classification, mixup takes two input series and a random mix coefficient λ, producing a blended series. This technique introduces smooth interpolations between time series instances and has been shown to improve robustness in sequence models. Mixup-based approaches have also been explored for sensor-based time series data, such as in health monitoring applications (Um et al., 2017).

*Magnitude Scaling and Warping*: This family of augmentations systematically alters the amplitude of a time series. A simple form is scaling, multiplying the entire series (or segments of it) by a constant factor. This mimics scenarios where the magnitude changes (e.g. due to sensor calibration differences or shifts in demand levels) while preserving the shape. More advanced methods such as *magnitude warping* apply different scale factors to different time points using interpolation between randomly sampled control points. This results in local stretching or compression of the signal amplitude. Such techniques allow models to generalize better across value ranges and have been shown effective in empirical comparisons of augmentation techniques (Iwana & Uchida, 2021).

*Synthetic Generative Methods (GANs, VAEs, Diffusion Models)*: These methods train a generative model to learn the data distribution and produce entirely new, plausible time series.

*TimeGAN* combines recurrent networks with adversarial training and an embedding loss to generate realistic synthetic sequences while preserving temporal dynamics. It has demonstrated strong performance in replicating both statistical and temporal properties of real-world time series (Yoon et al., 2019).

*Variational Autoencoders (VAEs)* on the other hand, offer a probabilistic approach to modeling time series, learning a latent space from which new sequences can be sampled. VAE-based augmentation has proven useful in domains where preserving underlying structure and interpretability is essential (Fortuin et al., 2019).

*Denoising Diffusion Models* represent a more recent and powerful class of generative models. These probabilistic models iteratively add and remove noise to synthesize data, capturing complex temporal dependencies. They have shown promise in producing high-fidelity time series that match intricate statistical properties like volatility clustering in finance or context dynamics in human activity (Tashiro et al., 2021).

## 4.2 Causal Modelling and Interpretability

*Disentangling Correlation and Causation:* A significant limitation in time series prediction arises from models confusing correlation with causation, leading to biased and potentially unjust predictions. Integrating causal inference frameworks, as recommended by Pearl and Mackenzie (2018), can help differentiate between mere correlations and genuine causal relationships, thereby enhancing predictive fairness and reliability. Causal models explicitly consider external interventions and structural changes, which allows them to remain valid across diverse contexts and avoid perpetuating historical biases (Pearl & Mackenzie, 2018).

*Enhanced Interpretability and Transparency:* Ensuring interpretability is fundamental to mitigating biases by making predictive processes understandable and transparent to stakeholders. Models with clear interpretability allow users and developers to identify how and why specific decisions are made, thus facilitating accountability and trust. (Burrell 2016, Coeckelbergh 2020) Fazelpour and Danks highlight the importance of transparency in algorithmic decision-making, arguing that interpretability is essential for aligning model outcomes with ethical and social values (Fazelpour & Danks 2021). Techniques such as SHAP (SHapley Additive exPlanations) and LIME (Local Interpretable Model-agnostic Explanations) can be leveraged to clarify model decisions, helping stakeholders critically evaluate predictive results and challenge potentially biased outcomes.

### 4.3 Ethical Design and Stakeholder Engagement

Ethical design in time series prediction systems demands a proactive and participatory approach, especially given the risks of algorithmic bias in critical domains like healthcare, energy, and public policy. (Brey and Dainow 2024) While many technical interventions aim to mitigate bias at the data or model level, truly ethical design must embed principles of fairness, accountability, and transparency throughout the system's entire lifecycle (Floridi & Cowls 2022, Tsamados et al. 2021).

Recent work underscores that fairness should not be reduced to a set of statistical constraints. Giovanola and Tiribelli argue that fairness is not only distributive but also socio-relational, grounded in a commitment to respect individuals as persons, not merely as data subjects. (Giovanola and Tiribelli 2022) They contend that fairness involves more than eliminating discrimination or achieving equal outcomes; it also includes fostering conditions that allow individuals to be acknowledged in their specificity, history, and vulnerability. Ethical design must therefore reflect these normative commitments from the outset, including in the way problems are framed and stakeholders are engaged.

Stakeholder engagement is crucial to this effort. Participatory design and co-development with affected communities enable the articulation of context-specific norms and values, which can serve as practical guides for ethical decision-making during system development. Heaton et al. note that trust and accountability in algorithmic systems are shaped not only by outcomes but by the perceived agency and responsibility embedded in system governance. (Heaton et al. 2023, Hanna et al. 2025) Therefore, engaging stakeholders meaningfully, especially those from vulnerable or marginalized groups, can help ensure that algorithmic tools serve their interests and uphold democratic legitimacy.

This participatory ethic also addresses issues of transparency and power asymmetry. The barriers to accountability described by Nissenbaum (1996), such as the problem of many hands and considering systems as scapegoat, are particularly pronounced in machine learning systems, where decision-making processes are complex, opaque, and distributed. (Nissenbaum 1996) Cooper et al. build on this insight, advocating for a relational model of accountability that clarifies who is responsible, for what, and under which circumstances.

(Cooper at al. 2022, Martin 2019) Such clarity requires institutional reforms, including algorithmic impact assessments and ongoing audits to formalize accountability pathways and validate model performance against fairness benchmarks.

Hacker proposes a complementary legal framework, integrating EU anti-discrimination and data protection laws to strengthen algorithmic fairness. He demonstrates how tools like Data Protection Impact Assessments (DPIAs) under the GDPR can be leveraged not only to audit algorithmic risk but to align system design with substantive equality standards. (Hacker 2018) This vision of "equal protection by design" reinforces ethical design with enforceable safeguards, helping prevent harms that might otherwise go unaddressed due to legal or institutional inertia.

Yet participatory ethics must go beyond formal compliance. As Matzner and Mann emphasize, emergent discrimination can arise from latent proxies, feedback loops, and structural inequities that are invisible to technical audits or legal review. For instance, models may reproduce discriminatory outcomes by inferring group membership indirectly (e.g., via postcode or consumption pattern), even when sensitive attributes are excluded. To address such phenomena, stakeholder engagement must include critical, intersectional, and post-colonial perspectives that challenge dominant narratives and reveal hidden harms.

As a consequence, ethical design in time series prediction systems should rest on three foundational pillars: 1. *Procedural Fairness*: Engage affected communities in setting model objectives, performance criteria, and fairness constraints through participatory and co-design methods. *Dynamic Accountability*: Build mechanisms for reflexive governance, such as audit trails, redress systems, and adaptive oversight that evolve in response to social feedback and model drift. *Translational Transparency*: Provide context-appropriate explanations of model logic and limitations tailored to different audiences, technical teams, regulators, and the general public.

This shows that ethical design and stakeholder engagement must be grounded in a broad understanding of fairness as both a moral and relational value. By integrating philosophical inquiry, legal safeguards, and participatory governance, we can better align time series prediction systems with democratic and inclusive social aims.

**4.4 Fairness Metrics and Validation Methods**

In time series prediction, fairness validation has historically relied on quantitative metrics derived from classification or regression tasks, e.g., demographic parity, equalized odds, or statistical parity difference. While these measures are useful for detecting and quantifying disparities across groups, their application to time series prediction is non-trivial. Temporal dependencies, non-stationarity, and context-specific risks introduce challenges that static fairness metrics often fail to address adequately. Standard fairness metrics often assume predefined, clearly demarcated protected groups. Yet, as Matzner and Mann argue, algorithmic profiling increasingly operates through proxies and latent correlations, leading to emergent forms of discrimination that are difficult to capture using traditional group-based metrics. (Matzner and Mann 2019) For example, time series models may use location-based features (e.g., zip code or usage patterns) as indirect predictors of race or socio-economic status. In these cases, group fairness metrics may underestimate real-world harms unless supplemented by qualitative and structural assessments.

To address this, fairness validation must be reframed through a multi-dimensional lens. Giovanola and Tiribelli advocate for a broader ethical conceptualization of fairness, one that extends beyond distributive concerns to include socio-relational respect and context sensitivity. (Giovanella and Tribelli 2021) Applying this perspective to validation implies not only comparing model outputs across demographic groups but also assessing how well the system aligns with the values and lived experiences of those affected. This calls for *context-aware validation strategies* that combine formal fairness metrics with participatory evaluation. Stakeholders, including domain experts, regulators, and community members, should be involved in defining what constitutes fair treatment in a given application. This reflects the position advocated by Hacker, who argues that technical fairness should be reinforced by enforceable legal safeguards, such as Data Protection Impact Assessments (DPIAs) under the GDPR and external fairness audits. (Hacker 2018) Validation should therefore incorporate not only statistical metrics, but also compliance with data protection principles and substantive equality standards.

Another consideration is what can be labeled *temporal fairness*: given that time series models operate over changing distributions, fairness must be evaluated over time. A model may appear fair at one point but drift into discriminatory behavior due to data shifts or policy changes. This supports calls for dynamic validation frameworks that monitor fairness continuously rather than at deployment alone. Audit trails, logging mechanisms, and fairness dashboards should be built into the system, allowing real-time alerts when fairness thresholds are breached. Additionally, *multi-metric approaches* should be favored over single-metric evaluations. (Hacker 2018, Birhane 2021) For instance, group fairness metrics may be combined with individual fairness measures (which assess consistency in treatment of similar individuals), causal fairness metrics (which evaluate whether sensitive attributes influence outcomes directly or indirectly), and counterfactual-based metrics (which simulate how decisions would differ under alternate group memberships). As Hacker points out, such complementary perspectives can help avoid over-reliance on narrow formalism and instead guide toward "equal treatment by design".

Validation should also consider downstream impact analysis, asking not only whether a model is fair in statistical terms, but also whether its deployment creates feedback loops that reinforce existing inequalities, as Matzner and Mann caution. (Matzner and Mann 2019) For example, an energy demand forecasting model that systematically underpredicts usage in low-income areas could contribute to under-provisioning and further entrench energy poverty. In such cases, fairness is not merely a matter of algorithmic outputs but of long-term socio-economic effects.

## 5. Conclusion

This paper has examined the ethical foundations and mitigation strategies for algorithmic bias in time series prediction, a domain that plays a crucial role in shaping outcomes across healthcare, energy forecasting, and economic planning. Drawing from interdisciplinary literature in computer science, philosophy, and social theory, we have shown that algorithmic bias is not a purely technical flaw but a systemic issue that arises across all stages of the algorithmic pipeline, from problem formulation and data collection to modeling choices and deployment contexts.

We have argued that fairness in predictive systems cannot be achieved solely through post hoc adjustments or performance tuning. Instead, ethical considerations must be embedded from the outset in the design and evaluation process. This includes adopting inclusive problem specifications, implementing fairness-sensitive metrics, engaging stakeholders meaningfully, and ensuring transparency and interpretability. The importance of accountability structures, both technical and institutional, has also been emphasized, especially in high-stakes applications where harm is often distributed unequally.

At the same time, we acknowledge the practical and institutional constraints that make the realization of these ethical ideals challenging. While this paper underscores the value of integrating fairness into algorithmic design, many practitioners still lack formal education in social justice and struggle to operationalize normative concepts in machine-readable form. Even when fairness-aware methods are proposed, they are often difficult to verify or adopt at scale, particularly in settings where research and development are driven by commercial or performance-based incentives. Social accountability mechanisms, in practice, tend to emerge only at later stages through government regulations or institutional oversight, rather than through the actions of individual developers. These realities highlight the need not only for technical innovation, but also for broader structural reforms, such as curricular changes, interdisciplinary collaboration, and supportive policy frameworks, that enable ethical intentions to translate into practical outcomes.

Achieving fairness in time series prediction is not a matter of choosing between accuracy and ethics. It requires an integrated approach that recognizes the co-dependence of technical and normative excellence. By embedding ethical reflection within algorithmic development, and by fostering environments that support socially responsible design, we can move toward predictive systems that are not only effective but also just.

Burrell, J. (2016). How the machine 'thinks': Understanding opacity in machine learning algorithms. *Big Data & Society, 3*(1).

Coeckelbergh, M. (2020). Artificial intelligence, responsibility attribution, and a relational justification of explainability. *Science and Engineering Ethics, 26*(4), 2051–2068.

Coeckelbergh, M. (2022). *The Political Philosophy of AI*. Polity Books.

Cooper, A. F., Robbins, R., & O'Leary, T. (2022). Accountability in an algorithmic society: Relationality, responsibility, and robustness in machine learning. In *Proceedings of the 2022 ACM Conference on Fairness, Accountability, and Transparency* (pp. 864–876).

Fazelpour, S., & Danks, D. (2021). Algorithmic bias: Senses, sources, solutions. *Philosophy Compass*. https://doi.org/10.1111/phc3.12760

Floridi, L., & Cowls, J. (2022). A unified framework of five principles for AI in society. In *Machine Learning and the City: Applications in Architecture and Urban Design* (pp. 535–545).

Fortuin, V., Hüser, M., Locatello, F., Strathmann, H., & Rätsch, G. (2018). SOM-VAE: Interpretable discrete representation learning on time series. *arXiv preprint arXiv:1806.02199*.

Giovanola, B., & Tiribelli, S. (2022). Beyond bias and discrimination: Redefining the AI ethics principle of fairness in healthcare machine learning algorithms. *AI and Society*. https://doi.org/10.1007/s00146-022-01431-z

Hacker, P. (2018). Teaching fairness to artificial intelligence: Existing and novel strategies against algorithmic discrimination under EU law. *Common Market Law Review, 55*, 1143–1186.

Hanna, M. G., Pantanowitz, L., Jackson, B., Palmer, O., Visweswaran, S., Pantanowitz, J., ... & Rashidi, H. H. (2025). Ethical and bias considerations in artificial intelligence/machine learning. *Modern Pathology, 38*(3), 100686.

Heaton, D., Hendriks, J., & Taddeo, M. (2023). The social impact of decision-making algorithms: Reviewing the influence of agency, responsibility and accountability on trust and blame. In *TAS '23: Proceedings of the First International Symposium on Trustworthy Autonomous Systems*, Article 11, 1–11.

Iwana, B. K., & Uchida, S. (2021). An empirical survey of data augmentation for time series classification with neural networks. *PLOS ONE, 16*(7), e0254841.
21

**Declaration: No Funding was involved**